\documentclass[aps,prb,twocolumn,superscriptaddress,longbibliography ]{revtex4-1}

\usepackage{graphicx}
\usepackage{amsmath}
\usepackage{amssymb}
\usepackage{amsfonts}
\usepackage{textcomp}

\newcommand{\Ea}{\ensuremath{{\cal E}_1}}
\newcommand{\Eb}{\ensuremath{{\cal E}_2}}

\newcommand{\Ed}{\ensuremath{{\cal E}_{1,2}}}

\newcommand{\ket}[1]{\left| #1\right>}
\newcommand{\bra}[1]{\left< #1\right|}

\begin{document}

\title{Exploring coherence of individual excitons in InAs quantum dots\\ embedded in natural photonic defects: influence of the excitation intensity}

\author{D.~Wigger}
\email[]{d.wigger@wwu.de}
\affiliation{Institut f\"{u}r Festk\"{o}rpertheorie, Universit\"{a}t M\"{u}nster, 48149 M\"{u}nster, Germany}

\author{V.~Delmonte}
\affiliation{Univ. Grenoble Alpes, F-38000 Grenoble, France}
\affiliation{CNRS, Institut N\'{e}el, "Nanophysique et semiconducteurs" group, F-38000 Grenoble, France}

\author{Q.~Mermillod}
\affiliation{Univ. Grenoble Alpes, F-38000 Grenoble, France}
\affiliation{CNRS, Institut N\'{e}el, "Nanophysique et semiconducteurs" group, F-38000 Grenoble, France}

\author{T.~Jakubczyk}
\email{tomasz.jakubczyk@neel.cnrs.fr}
\affiliation{Univ. Grenoble Alpes, F-38000 Grenoble, France}
\affiliation{CNRS, Institut N\'{e}el, "Nanophysique et semiconducteurs" group, F-38000 Grenoble, France}

\author{F.~Fras}
\affiliation{Univ. Grenoble Alpes, F-38000 Grenoble, France}
\affiliation{CNRS, Institut N\'{e}el, "Nanophysique et semiconducteurs" group, F-38000 Grenoble, France}
\affiliation{Universit\'{e} de Strasbourg, CNRS, IPCMS, UMR 7504, F-67000 Strasbourg, France}

\author{S.~Le-Denmat}
\affiliation{Univ. Grenoble Alpes, F-38000 Grenoble, France}
\affiliation{CNRS, Institut N\'{e}el, "Nanophysique et semiconducteurs" group, F-38000 Grenoble, France}

\author{D.~E.~Reiter}
\affiliation{Institut f\"{u}r Festk\"{o}rpertheorie, Universit\"{a}t M\"{u}nster, 48149 M\"{u}nster, Germany}

\author{S.~H\"{o}fling}
\affiliation{Technische Physik, University of W\"{u}rzburg, 97074 W\"{u}rzburg, Germany}

\author{M.~Kamp}
\affiliation{Technische Physik, University of W\"{u}rzburg, 97074 W\"{u}rzburg, Germany}

\author{G.~Nogues}
\affiliation{Univ. Grenoble Alpes, F-38000 Grenoble, France}
\affiliation{CNRS, Institut N\'{e}el, "Nanophysique et semiconducteurs" group, F-38000 Grenoble, France}

\author{C.~Schneider}
\affiliation{Technische Physik, University of W\"{u}rzburg, 97074 W\"{u}rzburg, Germany}

\author{T.~Kuhn}
\affiliation{Institut f\"{u}r Festk\"{o}rpertheorie, Universit\"{a}t M\"{u}nster, 48149 M\"{u}nster, Germany}

\author{J.~Kasprzak}
\email[]{jacek.kasprzak@neel.cnrs.fr}
\affiliation{Univ. Grenoble Alpes, F-38000 Grenoble, France}
\affiliation{CNRS, Institut N\'{e}el, "Nanophysique et semiconducteurs" group, F-38000 Grenoble, France}

%%%%%%%%%%%%%%%%%%%%%%%%%%%%%%%%%%%%%%%%%%%%%%%%%%%%%%%%%%%%%%
%%%%%%%%%%%%%%%%%%%%%%%%%%%%%%%%%%%%%%%%%%%%%%%%%%%%%%%%%%%%%%
%%%%%%%%%%%%%%%%%%%%%%%%%%%%%%%%%%%%%%%%%%%%%%%%%%%%%%%%%%%%%%

\begin{abstract}
The exact optical response of quantum few-level systems depends crucially on the exact choice of the incoming pulse areas. We use four-wave mixing (FWM) spectroscopy to infer the coherent response and dynamics of single InAs quantum dots (QDs) and study their pulse area dependence. By combining atomic force microscopy with FWM hyperspectral imaging, we show that the retrieved FWM signals originate from individual QDs enclosed in natural photonic defects. The optimized light-matter coupling in these defects allows us to perform our studies in a wide range of driving field amplitudes. When varying the pulse areas of the exciting laser pulses the so-called Rabi rotations can be resolved by the two-pulse FWM technique. We investigate these Rabi rotations within two- and three-level systems, both theoretically and experimentally, and explain their damping by the coupling to acoustic phonons. To highlight the importance of the pulse area influence, we show that the phonon-induced dephasing of QD excitons depends on the pulse intensity.
\end{abstract}

%%%%%%%%%%%%%%%%%%%%%%%%%%%%%%%%%%%%%%%%%%%%%%%%%%%%%%%%%%%%%%

\date{\today}

\maketitle

%%%%%%%%%%%%%%%%%%%%%%%%%%%%%%%%%%%%%%%%%%%%%%%%%%%%%%%%%%%%%%
\section{Introduction}
Measuring and manipulating the coherence of single emitters in semiconductor nanostructures, for example excitons in quantum dots (QDs), is a challenging research field within the optics of condensed matter systems. The ongoing progress in this field enhances the potential for implementing such solid state structures as qubits in information technologies exploiting optical interfaces.\cite{LodahlRMP15} To optimize the control of single emitters, the interfacing of the solid state system with light has to be controlled with high accuracy. Modern optical techniques allow to create light beams of almost arbitrary dynamics,\cite{weinerOC11} frequency modulation\cite{weiner2011} and even vortex beams with higher angular momenta.\cite{andrews2011} On the solid state side, the epitaxy of QDs and chemical synthesis of nanocrystals\cite{bimberg1999} has reached a tremendous quality. Additionally, the fundamental interaction between light and the single emitter has to be known in detail. Coherent nonlinear spectroscopy of QD excitons provides a powerful technique to study the interplay between the emitter and the optical fields, inferring both local and propagative phenomena.\cite{KasprzakNPho11,  AlbertNatComm13} In particular, by combining heterodyne detection with spectral interferometry one can access and control coherence dynamics in single QDs by investigating their four-wave mixing\cite{MermillodOptica16} (FWM) and six-wave mixing\cite{FrasNatPhot16, MermillodPRL16} responses.

Retrieving FWM signals from a single, strongly-confined exciton embedded in a planar sample is challenging\cite{KasprzakNJP13} as the signal co-exists with a resonant background, dominating FWM signals typically by 6-8 orders of magnitude in the field amplitude. We have recently shown that using nanophotonic devices with embedded QDs - specifically microcavities,\cite{FrasNatPhot16, MermillodOptica16} photonic-waveguide antennas\cite{MermillodPRL16} and deterministic micro-lenses\cite{jakubczykACSP16} - one can penetrate efficiently across the vacuum-dielectric boundary and intensify the electromagnetic field locally around the QD. This drastically decreases the inconvenient resonant background, improving the signal-collection efficiency considerably and allows for a variation of the driving intensities over a wide range. In a recent study, we have noticed that the dynamics of the FWM signals generated in single QDs strongly depends on the pulse areas acting on the exciton complexes.\cite{MermillodOptica16}

Having recognized the importance of the exact knowledge of the supplied pulse areas on the QD coherent response, in this paper, we focus on the intensity dependence of two-pulse FWM signals. We thereby analyze the Rabi rotations of individual excitons strongly-confined in InAs QDs. The damping of the rotations in a two-level system will be explained by the coupling to longitudinal acoustic (LA) phonons. We extend the study of Rabi rotations to a three-level system where the interplay between the transitions makes the situation more involved. In parallel, in Ref.\,[\onlinecite{jakubczykACSP16}] we have studied the influence of the exciton-phonon coupling on the FWM signal dynamics. We found that the creation of phonon wave packets leads to a loss of exciton coherence within a few picoseconds, which is called phonon-induced dephasing (PID). In this work, we will show that the PID effect also depends crucially on the pulse area of the driving laser pulses, when the pulse durations are in the range of $\tau\approx 0.5$~ps and therefore in the range of the typical timescale of the interacting phonons.

In the experiment we use a train of short pulses generated by a Ti:Sapphire laser, which is split into a pair with field amplitudes $\Ea$ and $\Eb$, and with a variable temporal delay $\tau_{12}$ (positive for $\Ea$ leading). Using acousto-optic modulators, the two optical beams are frequency shifted by radio-frequencies $\Omega_1=80$~MHz and $\Omega_2=80.77$~MHz, respectively. The FWM signal is selected in reflectance via interference with a reference field shifted by $2\Omega_2-\Omega_1$. This FWM heterodyne beating carries the third-order polarization $\Ea^{\star}\Eb^2$ and higher order terms with the same phase evolution, which dominate at high pulse areas. Details regarding the employed experimental configuration can be found in Ref.~[\onlinecite{FrasNatPhot16}].

The QDs are embedded in an epitaxially grown, one-lambda semiconductor microcavity, with the mirrors formed by distributed Bragg reflectors (DBR). We use the same sample as in Ref.~[\onlinecite{MermillodOptica16}] and~[\onlinecite{FrasNatPhot16}]. Its quality factor of 170 results in a cavity mode width of about 12~meV, well adapted to the spectral spread of femtosecond pulses. The mode centre can be tuned between 910~nm and 915~nm, owing to a slight gradient of the cavity width along the sample, in order to match the QD transition energies. An exceptionally high efficiency of the FWM signal retrieval has been recently reported for this structure,\cite{MermillodOptica16, FrasNatPhot16} as a result of two main factors. Firstly, the impinging excitation feeds the intra-cavity field, while avoiding spectral filtering, and is amplified at the QD location by the field cycling within the cavity. Secondly, as will be discussed in more detail in the next paragraph, the sample contains photonic defects\cite{zajacPRB12,MaierOptEx14} naturally forming during the microcavity growth, providing extra spatial localization for the optical excitation and extraction efficiency of up to 40\%.

\section{Imaging}

\begin{figure}
\includegraphics[width=\columnwidth]{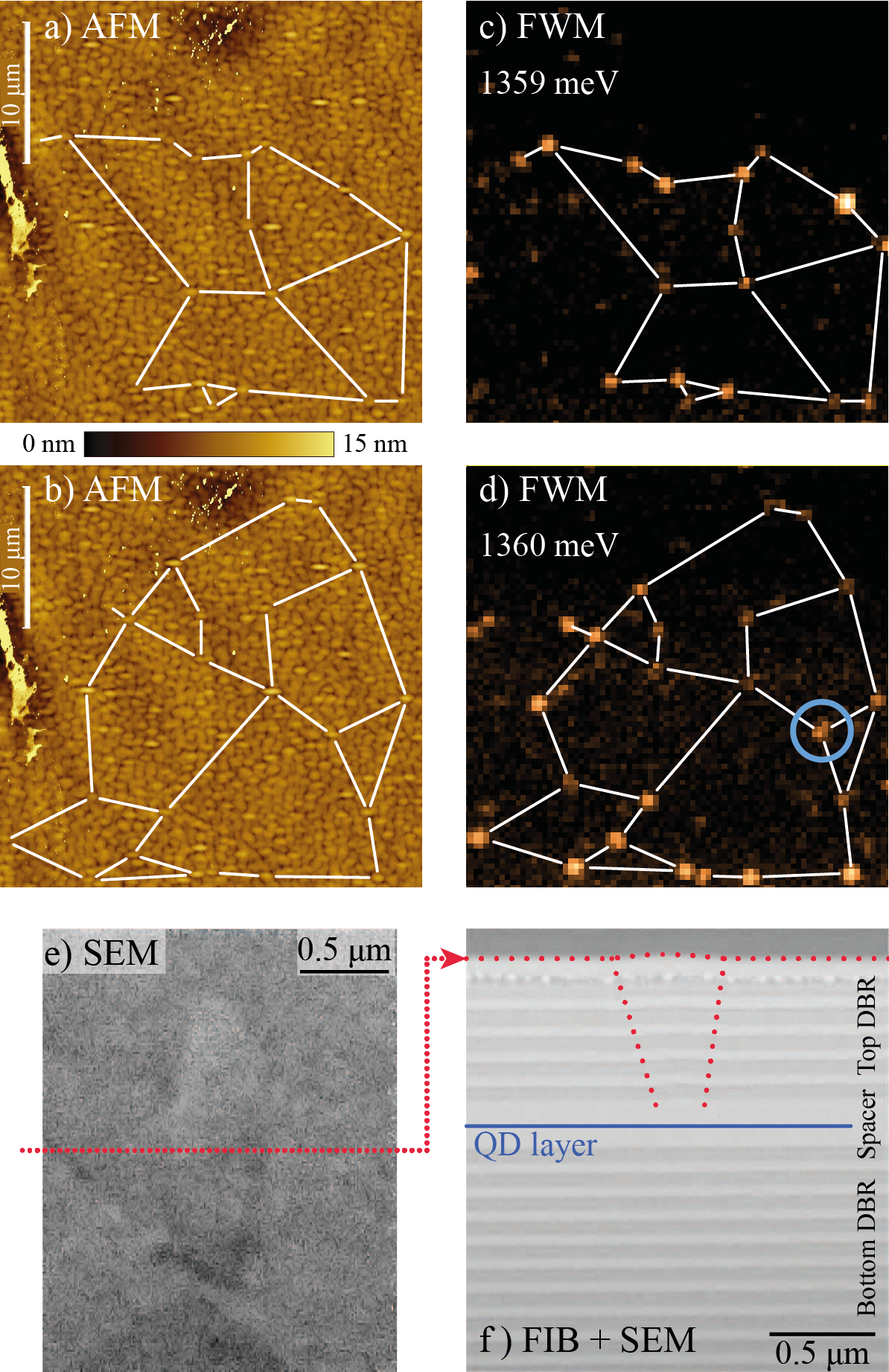}
\caption{
{\bf Photonic defects.}
a,b) Atomic-force microscopy carried out on the low-Q microcavity revealing natural photonic defect on the surface. c,d) Examples of FWM images for two different energies. The grids  illustrate apparent correlation between AFM and FWM. $\tau_{12}=0.2$~ps, $(P_1,\,P_2)=(200,\,400)$~nW. c) Scanning electron microscopy (SEM) image of one photonic defect. d) SEM image of the cross section of a photonic defect after cutting it with a focused ion beam (FIB) along the dotted line in c). The blue circle in d) marks the QD used for the FWM measurements in Sec.~\ref{sec:exp}.
\label{fig:defect}
}
\end{figure}

In Fig.~\ref{fig:defect}~a) and b) we show the sample topography recorded with atomic-force microscopy (AFM), revealing oval defects covering the whole surface. The close up scanning electron microscope (SEM) image of a single defect in e) provides a typical lateral size of a few microns and a height of 10 -- 15~nm. Their shape is elongated along the $\left[1\,\overline{1}\,0\right]$-direction. The distribution of the defects is random, with a characteristic spatial density of 0.1~\textmu m$^{-2}$. FWM hyperspectral imaging\cite{KasprzakNPho11, FrasNatPhot16, MermillodOptica16} was carried out on the same area as for the AFM and two exemplary spectral frames of this mapping are presented in Fig.~\ref{fig:defect}~c) and d) for photon energies of 1359~meV and 1360~meV, respectively. By comparing numerous AFM with FWM mappings [typical examples are shown in a)-d), where we compare a) with c) and b) with d)], it can be recognized that the spatially localized spots with intense FWM signals coincide with locations of photonic defects. As a guide to the eye in c) and d) we plot grids connecting bright FWM features. When overlaying these with the AFM images in a) and b), respectively, we find that the grid nodes coincide with distinct defects on the surface of the sample. Similar correlations were found when corroborating AFM topography with photoluminescence mapping of the same area (not shown). These observations indicate that the defects provide an enhanced in- and out-coupling efficiency and also suggest that the defects are self-aligned with the QDs.\cite{zajacPRB12,MaierOptEx14} A definite display of the latter would require QD imaging on a nanometer scale, for instance with transmission electron microscopy.

To elucidate the reason for the increased FWM signals at the positions of the photonic defects we have cut one of the defects with a focused ion beam (FIB) and scanned the cross section with an SEM. The result is given in Fig.~\ref{fig:defect}~f) and shows clearly the layered structure of the top and bottom DBR. The QD layer, marked by the blue line, is located in the center of the spacer between the two DBRs. The dotted red line at the top highlights the lens shaped contour of the defect at the surface. When we follow this pattern further down into the sample we find that the deformation is carried on into the layers of the top DBR toward the QD layer. This microscopic patterning of the photonic structure, in the form of a lens-like deformation conically propagating towards the surface (marked as dotted red line), leads to the efficient in- and out-coupling of the light,\cite{zajacPRB12,MaierOptEx14} as pointed out in simulations.\cite{DingPRB13} These defects allow us to tune the driving field amplitudes at the locations of the emitters over a large range. We have observed different spatial patterns of the FWM signal, when varying the powers $P_{1,2}$ of the pulses $\mathcal E_{1,2}$ (not shown). This is due to the distribution of optical coupling conditions realized by different defects, supplying different pulse areas $\theta_{1,2}$ at the QD layer, when scanning the sample surface.

\section{Theory}\label{sec:theory}

\begin{figure}
\centering
\includegraphics[width=0.9\columnwidth]{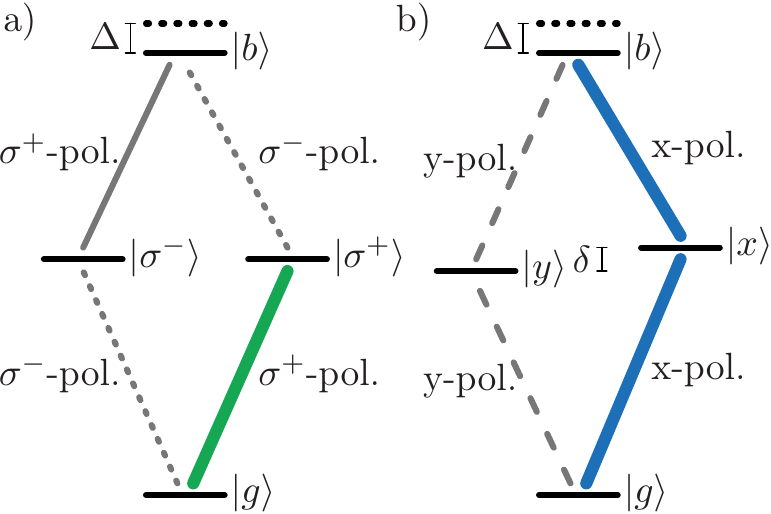}
\caption{
{\bf Schematic picture of the QD exciton transitions.}
a) Circularly polarized excitation is restricted to the two-level system g$\sigma^+$. b) Linearly x-polarized excitation reaches a three-level system (gxb).
\label{fig:scheme}
}
\end{figure}

Excitons in a QD provide an ideal playground to study different few-level schemes within the same quantum system. Restricting our study onto the timescale of a few picoseconds, we can choose between a two- or a three-level system by selecting circularly or linearly polarized light for the excitation, respectively. As depicted in Fig.~\ref{fig:scheme}~a), an excitation with right circularly polarized laser pulses drives the transition between the ground state $\ket{g}$ and the single exciton state $\ket{\sigma^+}$. To reach the biexciton state $\ket{b}$ a second pulse with left circular polarization would be needed. In general, the cylindrical symmetry of the QD is broken, which leads to the coupling between the single exciton states. As a result, the energy eigenstates are the linearly polarized states:
\begin{subequations}
\begin{eqnarray}
\ket{x} &=& \frac{1}{\sqrt{2}}\left(\ket{\sigma^+}+\ket{\sigma^-}\right)\ ,\\
\ket{y} &=& \frac{i}{\sqrt{2}}\left(\ket{\sigma^+}-\ket{\sigma^-}\right)\ .
\end{eqnarray}
\end{subequations}
The linearly x-polarized light is a superposition of $\sigma^+$ and $\sigma^-$ and therefore the biexciton state is naturally reached with a single pulse. This shows that the system has to be modeled by three levels, when driven with x-polarized pulses. This scenario is summarized in Fig.~\ref{fig:scheme}~b), where also the energy reduction $\Delta$ of $\ket{b}$ with respect to twice the single exciton energy is of importance, i.e.,
\begin{equation}
\Delta = 2\hbar\omega_x - \hbar\omega_b\ ,
\end{equation}
which is called the biexciton binding energy (BBE). $\Delta$ is usually on the order of a few meV.\cite{MermillodOptica16}

The coupling between the two $\sigma$-excitons further leads to an energy splitting between the linearly polarized excitons, known as fine structure splitting (FSS) $\delta$, which in our case is typically on the order of a few tens of micro-electronvolts.\cite{MermillodOptica16} One consequence of the FSS is that the coherences in the $\sigma$-polarized basis are not stationary, but transform into each other on the timescale of tens to hundreds of picoseconds.\cite{KasprzakPRB08, KasprzakNJP13, MermillodOptica16} Therefore it is also necessary to restrict the studies to times in the few picosecond range to certainly reduce the whole system to only two levels in the case of excitation by circularly polarized pulses.

\subsection{Two-level system}
We first focus on the instructive case of the two-level system under $\sigma^+$-polarized excitation. The Hamiltonian for the coupled exciton-light system in the usual rotating wave approximation is given by
\begin{equation}
H^{(2)} = \hbar\omega_{\sigma}\ket{\sigma^+}\bra{\sigma^+} - \sum_{i,j=1}^2 \hbar\mathcal M^{(2)} \ket{i}\bra{j}
\end{equation}
where the interaction matrix is given by
\begin{equation}
\mathcal M^{(2)} = \begin{pmatrix}
0 & M_2 \\
M_2^\ast & 0
\end{pmatrix}\ .
\end{equation}
The transition matrix element
\begin{equation}
M_2 = \sum_n \frac{\mu}{\hbar} e^{i \varphi_n} \mathcal E_n(t)=\sum_n \frac{1}{2} e^{i \varphi_n} \omega^{(n)}_{\rm R}(t)e^{-i\omega_{\rm L}t}
\label{eq:M_2L}
\end{equation}
is determined by a sequence of laser pulses with the pulse areas
\begin{equation}
\theta_n=\frac{2}{\hbar}\int \mu |\mathcal E_n(t)|{\rm d}t\ ,
\label{eq:theta_2L}
\end{equation}
the dipole moment $\mu$, relative phase factors with $\varphi_n$ and the instantaneous Rabi frequencies $\omega^{(n)}_{\rm R}(t)$, which describe the temporal envelopes of the pulses $\mathcal E_n$. The central frequency of the laser pulses is $\omega_{\rm L}$.

\begin{figure}
\includegraphics[width=\columnwidth]{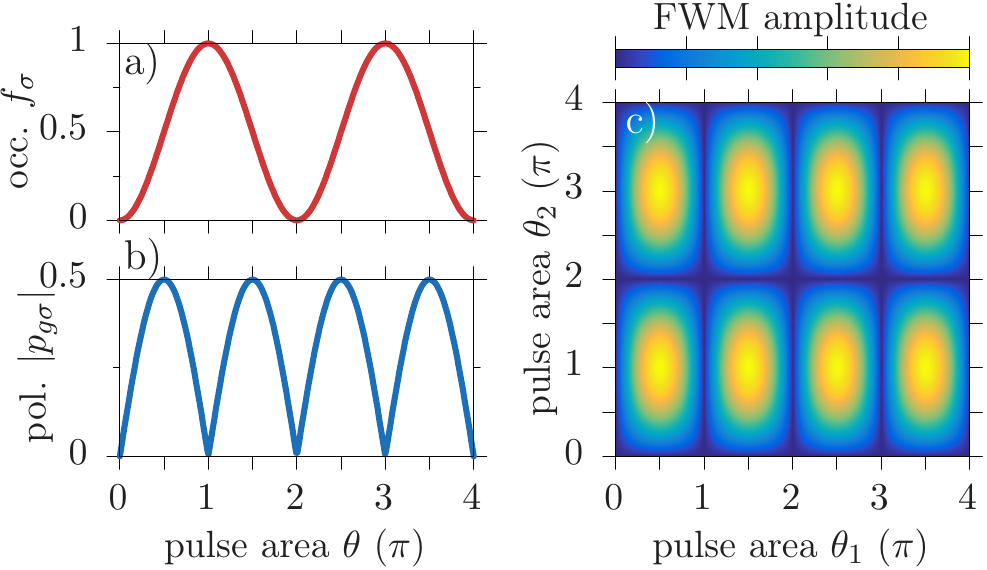}
\caption{
{\bf Excitation in the two-level system.}
a) Occupation of the single $\sigma^+$-exciton state as a function of pulse area. b) Absolute value of the microscopic polarization between $\ket{g}$ and $\ket{\sigma^+}$. c) FWM amplitude as a function of pulse areas $\theta_1$ and $\theta_2$ (2D Rabi rotations).
\label{fig:2level}
}
\end{figure}

When considering a single, resonant excitation, the final properties of the exciton state are unequivocally determined by the pulse area $\theta$. The final occupations of the two states $f_\sigma=\left<\ket{\sigma^+}\bra{\sigma^+}\right>$ and $f_g=\left<\ket{g}\bra{g}\right>$ and the microscopic polarization $p_{g\sigma}=\left<\ket{g}\bra{\sigma^+}\right>$ are given by
\begin{subequations}
\begin{eqnarray}
f_\sigma &=& \sin^2\left(\frac{\theta}{2}\right)\ ,\\
f_g &=& 1-f_\sigma\ , \\
p_{g\sigma} &=& \frac{i}{2}\sin\left(\theta\right)\ .
\end{eqnarray}
\end{subequations}
The exciton occupation $f_\sigma$ is plotted in Fig.~\ref{fig:2level}~a) and shows the periodic Rabi rotations, where a pulse area of $\theta=\pi$ fully inverts the system from $f_\sigma=0$ to $f_\sigma=1$. The polarization in b) accordingly vanishes when the system is either fully excited or de-excited and is maximal $\left|p_{g\sigma}\right|=0.5$ when the system is in an equal superposition of $\ket{g}$ and $\ket{\sigma^+}$.

The spectroscopy method to gain information about the polarization in the system is the FWM technique. To model the amplitude of the FWM signal $S_{\rm FWM}$ in the regime of ultrafast laser pulses we choose the pulse functions in Eq.~\eqref{eq:M_2L} in the delta-pulse limit, i.e.,
\begin{equation}
|\mathcal E_n(t)| = \frac{\hbar}{2\mu}\theta_n\delta\left(t-t_n\right)\ .
\end{equation}
To generate a FWM signal the system has to be excited by at least two pulses. While through a three-pulse excitation the occupations in the system can be analyzed,\cite{MermillodOptica16} we here restrict ourselves to the two-pulse scheme analyzing the coherences properties. The FWM-polarization after the two-pulse excitation is selected by a specific combination of the appearing phase factors $e^{i \varphi_n}$. The term proportional to the phase factor with $2\varphi_2-\varphi_1$, according to the choice of radio-frequency shifts $2\Omega_2-\Omega_1$ in the experiment, will give us the FWM-polarization $p_{\rm FWM}$. The detected signal amplitude will then be proportional to the absolute value of this polarization.

While the delay dynamics of the FWM signals is in detail investigated in Ref.~[\onlinecite{MermillodOptica16}], we are here interested in its dependence on the pulse areas $\theta_1$ and $\theta_2$ of the two driving pulses $\Ea$ and $\Eb$, which is given by
\begin{equation}
S_{\rm FWM}\sim |p_{\rm FWM}| = \left|\sin(\theta_1)\sin^2\left(\frac{\theta_2}{2}\right)\right|\ .
\label{eq:FWM2}
\end{equation}
This result for the FWM amplitude is shown in Fig.~\ref{fig:2level}~c) as a function of $\theta_1$ and $\theta_2$, we call this kind of plot 2D Rabi rotations. We find the strongest signal for odd multiples of $\theta_1 = \pi/2$ and $\theta_2 = \pi$, as found in pioneering FWM experiments on single quantum states.\cite{PattonPRL05} This can be easily understood when recalling the Bloch sphere representation of the exciton state. To get a strong signal the polarization after the first pulse has to be maximal, i.e., the Bloch vector has to point on the equator of the sphere, which is found for $\theta_1$ being odd multiples of $\pi/2$. The FWM signal is proportional to the FWM polarization after the second pulse, therefore the Bloch vector should still point to the equator after the second pulse. This is achieved with a half rotation of the Bloch vector, i.e., for pulse areas $\theta_2$ of odd multiples of $\pi$.

Within this basic two-level system the dynamics of the FWM signal is trivial, because without any coupling to phonons or any other type of dephasing it is just a constant function of the delay $\tau_{12}$ between the pulses.\cite{MermillodOptica16} Nevertheless, the Rabi rotations are important on the one hand to maximize the FWM signal and on the other hand to perform a calibration that correlates the pulse intensities in the experiment to pulse areas in the theory.  In Sec.~\ref{sec:exp} we will see that the details of the exciton-phonon coupling lead to a damping of the Rabi rotations, which makes the intuitive picture of the two-level system more involved.

\subsection{Three-level system}
\begin{figure}
\includegraphics[width=\columnwidth]{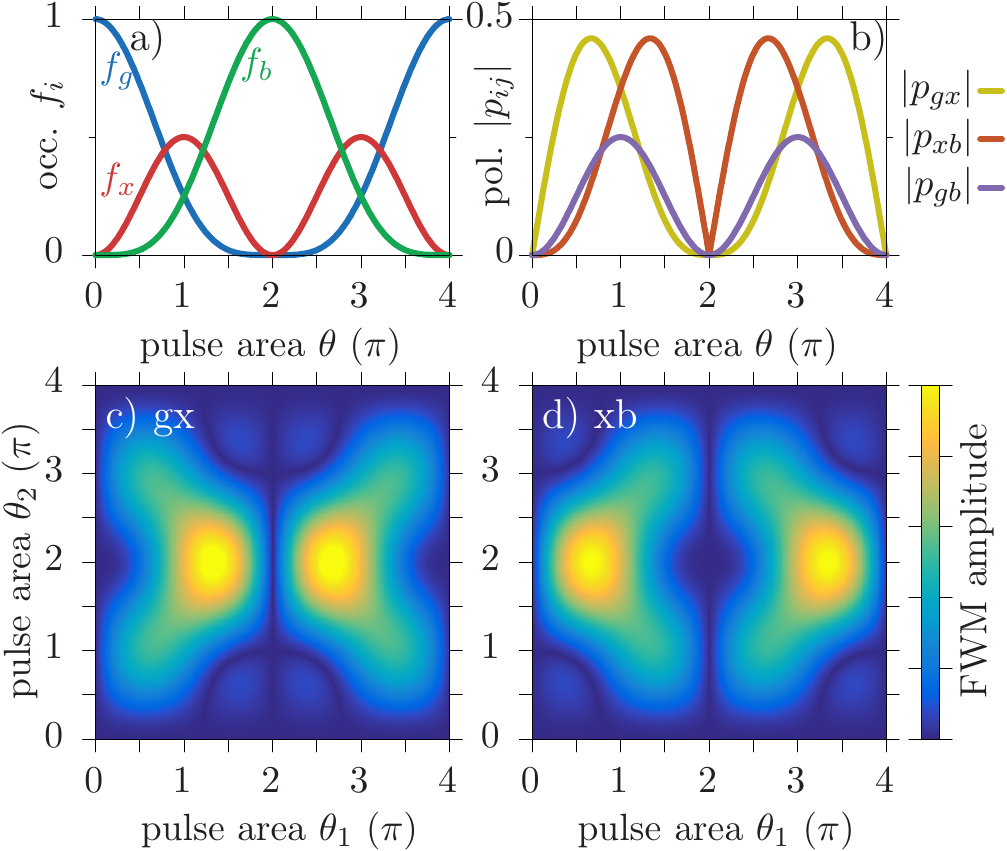}
\caption{
{\bf Excitation in the three-level system.}
a) Occupations of all three states ($f_g$, $f_x$ and $f_b$) as a function of pulse area. b) Absolute values of the polarizations ($\left|p_{gx}\right|$, $\left|p_{xb}\right|$ and $\left|p_{gb}\right|$). c,d) FWM amplitudes as a function of the pulse areas $\theta_1$ and $\theta_2$ for $\tau_{12}=0$. c) gx-transition, d) xb-transition.
\label{fig:3level}
}
\end{figure}

We increase the complexity and turn to the linearly x-polarized excitation sketched in Fig.~\ref{fig:scheme}~b). The Hamiltonian for the coupled exciton-light system is given by
\begin{equation}
H^{(3)} = \hbar\omega_{x}\ket{x}\bra{x} + (2\hbar\omega_{x}-\Delta)\ket{b}\bra{b} - \sum_{i,j=1}^3 \hbar\mathcal M^{(3)} \ket{i}\bra{j}\ ,
\end{equation}
We now assume that the spectral width of the driving laser pulses is larger than the BBE $\Delta$ and centered around the single exciton energy $\hbar\omega_x$, and that the ground state to biexciton transition is not driven. This is modeled by the transition matrix
\begin{equation}
\mathcal M^{(3)} = \begin{pmatrix}
0 & M_3 & 0\\
M_3^\ast & 0 & M_3\\
0 & M_3^\ast & 0
\end{pmatrix}\ ,
\end{equation}
where the matrix element is given by
\begin{equation}
M_3 =  \sum_n \frac{\mu}{\hbar} e^{i \varphi_n} \mathcal E_n(t)\sum_n \frac{1}{2\sqrt{2}} e^{i \varphi_n} \omega^{(n)}_{\rm R}(t)e^{-i\omega_{\rm L}t}
\label{eq:M_3L}
\end{equation}
and the pulse area reads
\begin{equation}
\theta_n=\frac{2\sqrt{2}}{\hbar}\int \mu |\mathcal E_n(t)|{\rm d}t\ .
\label{eq:theta_3L}
\end{equation}
At this point it is important to note, that compared to Eq.~\eqref{eq:theta_2L} the pulse area is scaled by an additional factor of $\sqrt{2}$. The reason for this are the final occupations and microscopic polarizations in the system, which are then given by
\begin{subequations}
\begin{eqnarray}
f_b &=& \sin^4\left(\frac{\theta}{4}\right)\ ,\\
f_x &=& \frac{1}{2} \sin^2\left(\frac{\theta}{2}\right)\ ,\\
f_g &=& \cos^4\left(\frac{\theta}{4}\right)\ , \\
p_{gx} &=& \frac{i}{\sqrt{2}}\sin\left(\frac{\theta}{2}\right)\cos^2\left(\frac{\theta}{4}\right)\ , \\
p_{xb} &=& \frac{i}{\sqrt{2}}\sin\left(\frac{\theta}{2}\right)\sin^2\left(\frac{\theta}{4}\right)\ , \\
p_{gb} &=& -\frac{1}{4}\sin^2\left(\frac{\theta}{2}\right)\ .
\end{eqnarray}
\end{subequations}
Due to this definition of the pulse area the maximum of the exciton occupation $f_x$ is found for a pulse area of $\theta=\pi$, as in the two-level system. All three occupations are shown in Fig.~\ref{fig:3level}~a) as a function of pulse area. We find that the maximal single exciton occupation only reaches $f_x=0.5$. This shows that the system cannot be prepared purely in $\ket{x}$ by a single pulse. But for a $2\pi$ pulse the system is fully inverted, i.e., it is purely in $\ket{b}$. Overall, the system has a periodicity of the Rabi rotations of $4\pi$. When focussing on the microscopic polarizations in b) we see that the optically inactive polarization $p_{gb}$ is only about half as strong as the two which directly couple to the optical field, namely $p_{gx}$ and $p_{xb}$.

To determine the FWM signals in the three-levels system for $\delta$-pulses we follow the same strategy as described above. Now the two optical active polarizations $p_{gx}$ and $p_{xb}$ lead to FWM signals. The amplitudes are given by $S^{\rm gx}_{\rm FWM}\sim \left|p_{\rm FWM}^{\rm gx}\right|$ and $S^{\rm xb}_{\rm FWM}\sim \left|p_{\rm FWM}^{\rm xb}\right|$ with
\begin{subequations}
\begin{eqnarray}
p_{\rm FWM}^{\rm gx} &= & \sin\left(\frac{\theta_1}{2}\right)\cos^2\left(\frac{\theta_1}{4}\right)\cos^2\left(\frac{\theta_2}{2}\right)\\
&-&2\sin\left(\frac{\theta_1}{2}\right)\sin^2\left(\frac{\theta_1}{4}\right) \cos\left(\frac{\theta_2}{2}\right)\sin^2\left(\frac{\theta_2}{4}\right) \notag\\
&&\hspace{4.5cm} \times \exp\left(i\frac{\Delta}{\hbar}\tau_{12}\right)\ , \notag \\
p_{\rm FWM}^{\rm xb} & = & \sin\left(\frac{\theta_1}{2}\right)\sin^2\left(\frac{\theta_1}{4}\right)\cos^2\left(\frac{\theta_2}{2}\right)\\
&-&2\sin\left(\frac{\theta_1}{2}\right)\cos^2\left(\frac{\theta_1}{4}\right) \cos\left(\frac{\theta_2}{2}\right)\sin^2\left(\frac{\theta_2}{4}\right) \notag\\
&&\hspace{4.2cm} \times \exp\left(-i\frac{\Delta}{\hbar}\tau_{12}\right)\ . \notag
\end{eqnarray}
\label{eq:FWM_3}\end{subequations}
Due to the mixing of the two polarizations into both FWM signals we find a quite involved dependence on the pulse areas $\theta_1$ and $\theta_2$ and an influence of the delay $\tau_{12}$ between the two exciting pulses. This periodicity of the delay is determined by the BBE $\Delta$ because the relative phase between the $p_{gx}$ and $p_{xb}$ rotates with their energy difference, i.e., with $\Delta$. Note that the commonly employed representation of Rabi rotations, as evolution of a state vector on a Bloch sphere, does not hold here anymore.

Figure~\ref{fig:3level}~c,d) shows the 2D Rabi rotations of the FWM amplitudes for the gx- and the xb-transition, respectively. For these pictures the pulse delay was chosen to $\tau_{12}=0$. We find that the strength of the FWM signal strongly depends on the pulse areas of both pulses and does not simply recover the rotations of the polarizations for a single pulse. The strongest signals are achieved for $\theta_2=2\pi$, which in analogy to the two-level system corresponds to a fully inverting pulse. But depending on the first pulse area the maxima of $S^{\rm gx}_{\rm FWM}$ and $S^{\rm xb}_{\rm FWM}$ do not coincide. The first one for $S^{\rm gx}_{\rm FWM}$ (c) is found for $\theta_1>\pi$, while the first one for $S^{\rm xb}_{\rm FWM}$ (d) appears at $\theta_1<\pi$. For the special case of multiples of $\theta_1=2\pi$  all FWM signals vanish, which is in agreement with the vanishing polarizations for this pulse area in b).

We see that already the consideration of a third level makes the Rabi rotations much more complex. Here it is not possible to optimize both of the FWM signals at the same time. The dynamics of the FWM in a three level system is often dominated by quantum beats which arise due to the different transition energies in the system. In Ref.~[\onlinecite{MermillodOptica16}] it was shown that the appearance of these beats strongly depends on the chosen pulse areas, which already shows the importance of an exact knowledge and calibration of the pulse areas in the experiment.

\begin{figure}
\includegraphics[width=\columnwidth]{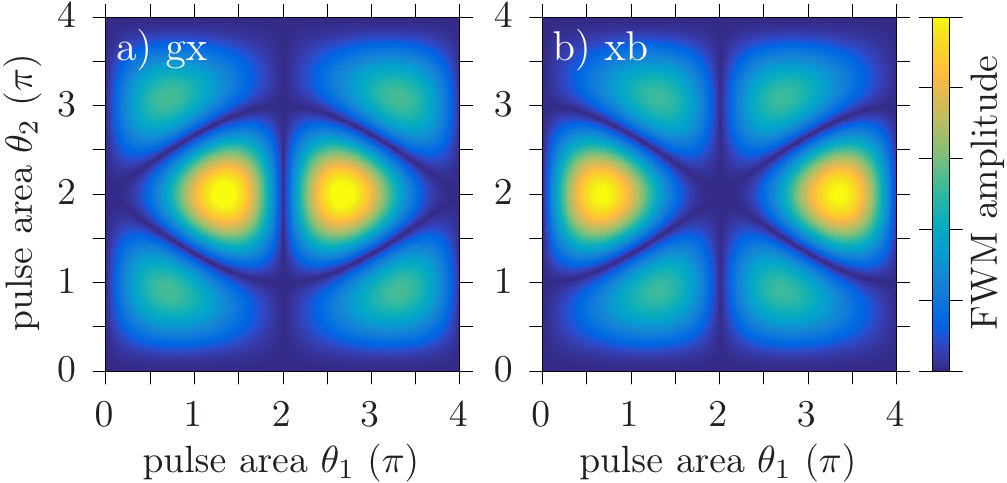}
\caption{
{\bf FWM for larger delay.}
a,b) Same as Fig.~\ref{fig:3level}~c,d) but for $\tau_{12}=\pi\hbar/\Delta$. \label{fig:3level_Delta_2}
}
\end{figure}

Details of the 2D Rabi rotation structure change significantly when increasing the delay to half a BBE period $\tau_{12}=\pi\hbar/\Delta$ as can be seen in Fig.~\ref{fig:3level_Delta_2}. The maxima stay approximately at the same positions, but for smaller and larger $\theta_2$ the picture looks quite different. This demonstrates that not only the specific choice of the pulse areas is crucial for the FWM signals but also the exact knowledge of the pulse delay is important. Animations for the complete time-dependence of the 2D Rabi rotations between Figs.~\ref{fig:3level} and ~\ref{fig:3level_Delta_2}, showing the smooth transition between the pictures, can be found in the Supporting Information.

Assuming a positive delay, i.e., pulse 2 arrives after pulse 1, in the three-level system it is also possible to retrieve FWM signals for the phase combination $2\varphi_1-\varphi_2$, which is not possible in two-level systems. In the phase combination $2\varphi_2-\varphi_1$ this corresponds to negative delays $\tau_{12}<0$ for which $\mathcal E_2$ arrives before $\mathcal E_1$. In this case the two FWM  signals are the same and they are proportional to the polarization between the ground state and the biexciton state after the first pulse, i.e., $S^{\rm gx}_{\rm FWM,\tau<0}=S^{\rm xb}_{\rm FWM,\tau<0}\sim\left|p^{\rm gx/xb}_{\rm FWM,\tau<0}\right|\sim\left|p^{(1)}_{gb}\right|$. Therefore the signal is called two-photon coherence. The pulse area dependence is given by
\begin{equation}
p^{\rm gx/xb}_{\rm FWM,\tau<0} = \sin^2\left(\frac{\theta_1}{2}\right) \sin\left(\frac{\theta_2}{2}\right) \cos^2\left(\frac{\theta_2}{4}\right)\ .
\label{eq:TPC}
\end{equation}
It is interesting to note that the signals are $2\pi$-periodic depending on $\theta_1$, while they are $4\pi$-periodic for positive delays. Another striking difference is that the signals vanish for $\theta_2=2\pi$, where they were maximal for positive $\tau_{12}$.

%%%%%%%%%%%%%%%%%%%%%%%%%
\subsection{Phonon coupling in a two-level system}
The main mechanism that leads to the dephasing of the exciton state on a picosecond timescale is the coupling to longitudinal acoustic phonons,\cite{forstnerPRL03,zrenner2002coh,krugelAPB05,ramsayPRL10,ramsayPRL10_2} where the dominant interaction in InGaAs-based QDs is the deformation potential coupling.\cite{krummheuerPRB05} The Hamiltonian of the two-level system including the coupling to phonons reads
\begin{eqnarray}
H^{(2)}_{\rm ph} &=& \hbar\omega_{\sigma}\ket{\sigma^+}\bra{\sigma^+} - \sum_{i,j=1}^2 \hbar\mathcal M^{(2)} \ket{i}\bra{j}\\
&+& \underbrace{\sum_{\bf q}\hbar g_{\bf q}\left(b_{\bf q}^{ }+b_{\bf q}^\dag\right)\ket{\sigma^+}\bra{\sigma^+}}_{H_{\rm x-ph}} + \sum_{\bf q}\hbar \omega_{\bf q} b_{\bf q}^\dag b_{\bf q}^{ }\ ,\notag
\end{eqnarray}
where $b_{\bf q}^\dag$ ($b_{\bf q}^{ }$) are the phonon creation (annihilation) operators with wave vector ${\bf q}$ and a linear dispersion $\omega_{\bf q}=c_{\rm LA}q$ is assumed. The phonon energies are in the range of a few meV and cannot lead to transitions between the exciton states. Therefore the interaction is modeled via the pure dephasing coupling between the exciton state $\ket{\sigma^+}$ and the phonons. The related coupling constant reads\cite{krummheuerPRB02}
\begin{equation}
g_{\bf q} = \sqrt{\frac{q}{2\rho c_{\rm LA} \hbar V}}\left[ D_{\rm e}e^{-\left(qa_{\rm e}\right)^2/4} - D_{\rm h}e^{-\left(qa_{\rm h}\right)^2/4} \right ]\ .
\end{equation}
On the one hand it depends on the semiconductor material via the mass density $\rho$, the sound velocity $c_{\rm LA}$ and the deformation potential coupling constants $D_{\rm e/h}$ for electron/hole and on the other hand on the QD geometry via the electron and hole localization lengths $a_{\rm e/h}$. $V$ is a normalization volume. From $g_{\bf q}$ we can calculate the phonon spectral density
\begin{equation}
J(\omega_{\rm ph}) = \sum_{\bf q} |g_{\bf q}|^2\delta(\omega_{\rm ph}-\omega_{\bf q})\ ,
\end{equation}
which is a measure for the exciton-phonon coupling strength depending on the phonon energy $\hbar\omega_{\rm ph}$.

\begin{figure}
\includegraphics[width=0.85\columnwidth]{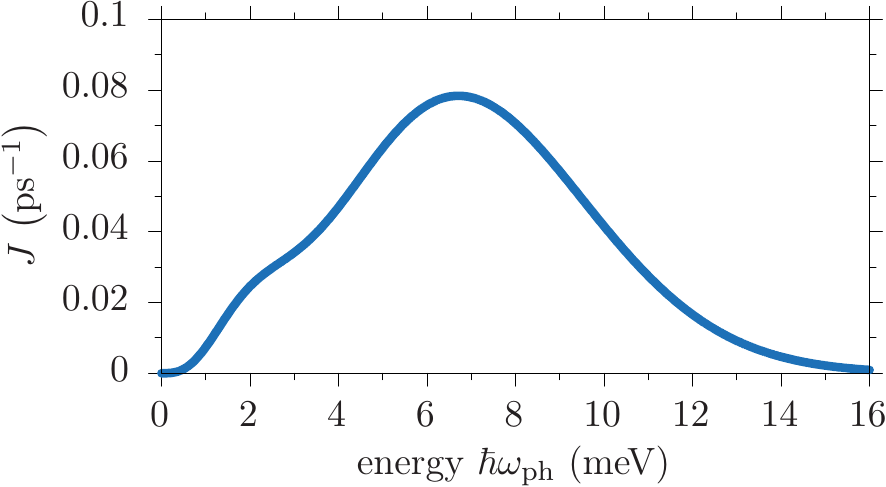}
\caption{
{\bf Phonon spectral density.}
Spectral density $J$ of the coupled phonons as a function of the phonon energy $\hbar\omega_{\rm ph}$ for electron and hole localization length $a_{\rm e}=8$~nm and $a_{\rm h}=2$~nm, respectively.
\label{fig:spec_dens}
}
\end{figure}

Figure~\ref{fig:spec_dens} shows the phonon spectral density for a QD with localization lengths $a_{\rm e}=8$~nm and $a_{\rm h}=2$~nm and deformation potentials $D_{\rm e}=7$~eV and $D_{\rm h}=-3.5$~eV,\cite{krummheuerPRB02} which will be used in Sec.~\ref{sec:PID}. Note that due to numerical reasons we here assume a spherical QD but can mimic a lens shaped dot by choosing quite different values for $a_{\rm e}$ and $a_{\rm h}$. We see that the energies of the coupled phonons basically spread between 1~meV and 14~meV. The coupling becomes ineffective for very small energies and energies above 14~meV.

The influence of the exciton-phonon coupling on the dephasing can best be understood in the dressed state basis, which is given by the eigenstates of the coupled exciton-light system.\cite{tannor2007} When the  light field is in resonance with the exciton transition the dressed states are given by
\begin{equation}
\ket{\psi^\pm} = \frac{1}{\sqrt{2}} \left(\ket{g} \mp \ket{\sigma^+}\right)
\end{equation}
with the energies
\begin{equation}
E^{\pm} = \frac12\hbar \left(-\omega_{\rm pol} \pm \omega_{\rm R}\right) \ ,
\end{equation}
where $\hbar\omega_{\rm pol}=\sum_{\bf q}|g_{\bf q}|^2/\omega_{\bf q}$ is the polaron energy. The energy splitting between the two dressed states \linebreak$E^+-E^-=\hbar\omega_{\rm R}$ is therefore determined by the Rabi frequency. Keeping in mind that the Rabi frequency is proportional to the amplitude of the driving laser fields, this is also the case for the splitting of the dressed states.

Transformed into the dressed state basis the exciton-phonon coupling Hamiltonian reads
\begin{eqnarray}
H_{\rm x-ph} &=& \sum_{\bf q} \frac{\hbar}{2}g_{\bf q}\left(b_{\bf q}^{ } + b^\dag_{\bf q}\right)\big(\ket{\psi^+}\bra{\psi^+}+\ket{\psi^-}\bra{\psi^-}\notag \\
&&\qquad\qquad -\ket{\psi^+}\bra{\psi^-}-\ket{\psi^-}\bra{\psi^+} \big)\ ,
\end{eqnarray}
where now transitions between the dressed states are mediated via the emission or absorption of a phonon given by $b^\dag_{\rm q}\ket{\psi^-}\bra{\psi^+}$ and $b^{ }_{\rm q}\ket{\psi^+}\bra{\psi^-}$, respectively. In analogy to the rotating wave approximation in the Jaynes-Cummings model we can neglect the processes that emit a phonon and excite the system or absorb a phonon and de-excite it, i.e., $b^\dag_{\rm q}\ket{\psi^+}\bra{\psi^-}$ and $b^{ }_{\rm q}\ket{\psi^-}\bra{\psi^+}$, in the discussion. Because this study is restricted to temperatures of $T\approx 5$~K we can also assume that only a small amount of phonons is thermally occupied and the phonon coupling is dominated by a relaxation from the upper dressed state to the lower one by the emission of phonons.\cite{ramsayPRL10,glasslPRB11,lukerPRB12,WiggerJPC14}

Together with the phonon spectral density we can conclude that the phonon effect on the exciton system will strongly depend on the laser intensity. When the splitting of the dressed states, which is governed by the laser intensity, agrees with large values of the spectral density $J$ the interaction is stronger than for very small and large splittings, i.e., for small and large laser intensities.\cite{WiggerJPC14}

To calculate the exciton polarizations in the phonon coupled system, from which we retrieve the FWM signals, we use a correlation expansion and truncate the equations of motion on the second order. The full set of equations can be found in Ref.~[\onlinecite{krugelPRB06}]. The FWM signal is then calculated numerically via a Fourier expansion with respect to the phases $\varphi_1$ and $\varphi_2$.\cite{haug2008}

%%%%%%%%%%%%%%%%%%%%%%%%%
\section{FWM intensity dependence in an exciton-biexciton system}
\label{sec:exp}
As highlighted in the Sec.~\ref{sec:theory} the pulse area is a decisive parameter when performing FWM spectroscopy of exciton complexes, as it governs not only the signal amplitudes, but also dynamics and couplings.\cite{MermillodOptica16} We assume that the system is driven with Gaussian pulses of the form
\begin{equation}
|\mathcal E_n(t)| = \frac{\hbar}{2\mu}\omega^{(n)}_{\rm R}(t) = \overline{\mathcal E}_n \exp\left[-\frac12 \left(\frac{t-t_n}{\tau}\right)^2\right]\ ,
\end{equation}
where $\overline{\mathcal E}_n$ is the amplitude and $\tau$ the duration of a pulse arriving at $t_n$. Then the power of the pulse is $P_n=|\overline{\mathcal E}_n|^2$, which leads to a pulse areas with
\begin{equation}
\theta_n \sim \tau\, \sqrt{P_n}\ .
\label{eq:theta_tau}
\end{equation}
Note that this pulse duration corresponds to a full width at half maximum of $2\sqrt{2\ln(2)}\,\tau$.

\subsection{Two-level system}\label{sec:exp_2}
When we apply the co-circular excitation, as described in Sec.~\ref{sec:theory}, we do not excite the biexciton state. When we further restrict our investigations to the few picosecond timescale, the other circularly polarized exciton does not contribute to the signals, such that we are dealing with a two-level system. Within this system the ordinary Rabi rotations should appear, which provide a calibration for the pulse areas.

\begin{figure}
\includegraphics[width=\columnwidth]{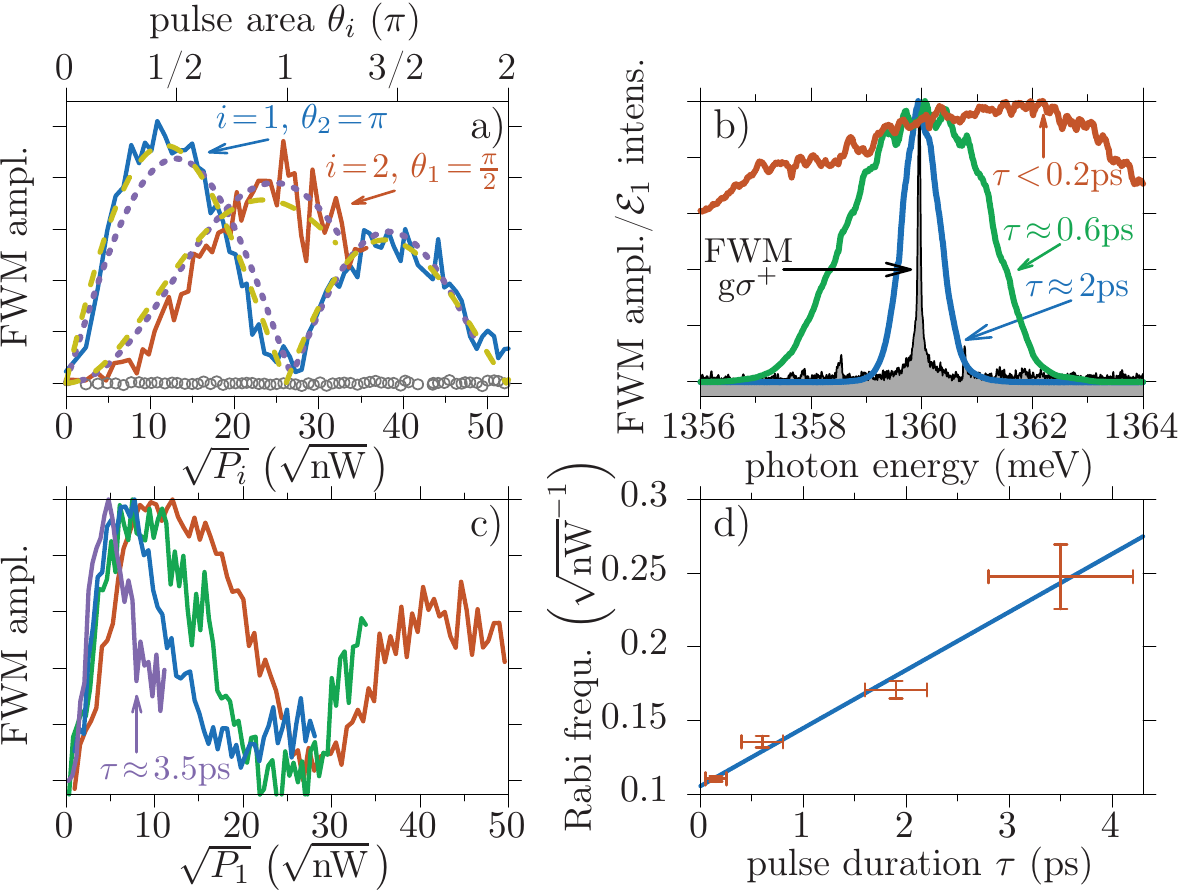}
\caption{
{\bf Rabi rotations in a two-level exciton system.}
a)~Spectrally-integrated FWM amplitude as a function of $\theta_1$ and $\theta_2$ showing Rabi rotations. Measurement in blue and orange for $\tau_{12}=0.2$~ps, theoretical fits from Eq.~\eqref{eq:FWM2} in dashed yellow. Full theory including exciton-phonon coupling in dotted violet for $\tau_{12}=0.1$~ps and $\tau=200$~fs pulses. b)~Spectral shape of the exciting pulses with different durations $\tau$ used in c) and FWM amplitude of targeted g$\sigma^+$-transition (filled area). c)~FWM amplitude as a function of $\sqrt{P_1}$ for different pulse durations $\tau$. Co-circular polarization of $\Ed$ and $\tau_{12}=0.2$~ps for panels a)-c). d) Rabi frequency retrieved from c) against pulse duration from b) with linear fit.
\label{fig:2level_exp}
}
\end{figure}

In Fig.~\ref{fig:2level_exp}~a) we present measurements of the Rabi rotations, when probing the FWM as a function of $\theta_1$ or $\theta_2$ and fixing the respective other. We have chosen the ground state to exciton transition (g$\sigma^+$) in the QD located in the defect marked by the blue circle in Fig.~\ref{fig:defect}~d). We see that the measured Rabi rotations are damped for growing pulse areas. In traditional Rabi rotation measurements, where the occupation (not the coherence as in our FWM study) after a single pulse excitation is measured as a function of the pulse area, the damping of the rotations happens due to the coupling to phonons.\cite{zrenner2002coh,krugelAPB05,ramsayPRL10,ramsayPRL10_2} As explained in Sec.~\ref{sec:theory} the exciton-phonon coupling leads to transitions between the two dressed states. For small laser intensities the splitting of the dressed states is small and corresponds to minor values of the phonon spectral densities. Therefore the interaction between the exciton and the phonons is negligible and the Rabi rotations follow the unperturbed hamonic prediction. When the pulse intensities and thereby the pulse areas increase also the dressed state splitting grows and gets into the range of larger values of the phonon spectral density. This leads to a rising influence of the phonons, which causes stronger dephasing. This dephasing directly leads to a decrease of the exciton occupation. In the full picture of the Rabi rotations one finds the fading of the measured signal with increasing pulse areas.\cite{ramsayPRL10} When increasing the pulse intensities even further the splitting of the dressed states would reach energies that are larger than the energy of the maximal exciton-phonon coupling in Fig.~\ref{fig:spec_dens}. From this point on the dephasing influence of the phonons would shrink again and the Rabi rotations should recover, which is called reappearance of the Rabi rotations in the literature.\cite{vagovPRL07,ramsayPRL10} While this regime cannot be reached easily with the laser pulses used in our approach, it was recently realized by using excitations with chirped laser pulses.\cite{kaldeweyARXIV17}

However for excitations with ultrafast laser pulses with durations $\tau\lesssim 200$~fs the measured damping of the FWM amplitude in Fig.~\ref{fig:2level_exp}~a) (blue and orange) is too strong to happen only due to phonon assisted transitions between the dressed states. In single pulse Rabi rotations at low temperatures multiple flops could be resolved.\cite{ramsayPRL10} In our case the strongest contribution to the discovered significant damping lies in the choice of the short delay $\tau_{12}=0.2$~ps between the two pulses. It is known that the FWM amplitude performs the characteristic PID drop (see Fig.~\ref{fig:PID}) for $\tau_{12}>0$ because of the coherence loss accompanying the emission of a phonon wave packet.\cite{jakubczykACSP16} The dependence of this effect on the driving pulse areas will be discussed in Sec.~\ref{sec:PID}. Because the FWM signal is absent before the second laser pulse and the described drop within a few ps, the FWM amplitude has a maximum at delays $\tau_{12}<1$~ps. We will see in Sec.~\ref{sec:PID} that this maximum of the FWM amplitude shifts in delay $\tau_{12}$ as a function of the pulse area. Typically in the measurement a short delay is chosen  in oder to minimize the dephasing between the pulses and retrieve a strong FWM signal. The delay is therefore in the range of the FWM maximum and the amplitude for varying pulse areas is affected by the slight changes of the FWM dynamics, i.e., the position of the maximum.

Note that this complex dependence of the exciton-phonon coupling on the laser pulse intensity is not covered by models that treat the interaction by a single phenomenological dephasing time $T_2$. The Rabi rotations retrieved from the full theory including the exciton-phonon coupling is given as dotted violet curve for the QD size considered in Fig.~\ref{fig:spec_dens} and Sec.~\ref{sec:PID}, pulse durations of $\tau=200$~fs and for a delay of $\tau_{12}=0.1$~ps. We see that the damping of the FWM amplitudes is well reproduced by the model. We also find that the decay can be simulated by simply adding an exponential damping of the FWM amplitude as function of the pulse area to Eq.~\eqref{eq:FWM2}. The fits in the plot (dashed yellow) still confirm the predicted pulse area dependences satisfyingly. It is quite remarkable that for this range of small pulse areas the complex interplay between the exciton and the phonons and the specific choice of the pulse delay can be modeled by a single exponential laser intensity dependence.

Coming back to the initial motivation for this measurement, it serves as a calibration of the pulse area for the following experiments in the three-level system. Specifically, we find that:
\begin{equation}
\theta = \pi/2\quad \widehat{=}\quad \sqrt{P} \approx 13.1\ \sqrt{\rm nW}\ .
\label{eq:calibration}
\end{equation}

So far, we have used pulse durations in the range of $\tau\lesssim 200$~fs to excite the system. For the investigation of the PID in Sec.~\ref{sec:PID} we will use longer laser pulses. Therefore we briefly wish to characterize also the Rabi rotations for increased pulse durations. Between $\tau=0.4$~ps and 0.8~ps we shape the pulses spectrally as shown in Fig.~\ref{fig:2level_exp}~b), which also presents the FWM signal of the g$\sigma^+$-transition. For durations above 1~ps we employ a pico-second Ti:Sapphire laser, instead of a femto-second one. In Fig.~\ref{fig:2level_exp}~c) we now show the FWM amplitude as a function of $\sqrt{P_1}$ for the different pulse durations $\tau$. For clarity, in Fig.~\ref{fig:2level_exp}~b) the narrow lineshape of the excitation pulse yielding the violet curve in c) ($\tau\approx 3.5$~ps) is not presented. We directly see that the Rabi frequency grows with increasing pulse duration, such that less pulse power $P_1$ is needed to attain $\theta_1=\pi/2$. To get a more quantitative picture we fit the Rabi rotations with the expected $|\sin(\theta_1)|$ dependence and extract the Rabi frequency. Figure~\ref{fig:2level_exp}~d) summarizes the fitted Rabi frequencies as a function of the pulse duration $\tau$, which was extracted from the Fourier transforms of the pulse spectra in b). Note that the duration of the longest pulse was estimated to $\tau=(3.5\pm 0.7)$~ps. The points confirm the linear relation between the pulse area and the pulse duration from Eq.~\eqref{eq:theta_tau} as can be seen from the linear fit in blue.

\subsection{Three-level system}

\begin{figure}
\includegraphics[width=\columnwidth]{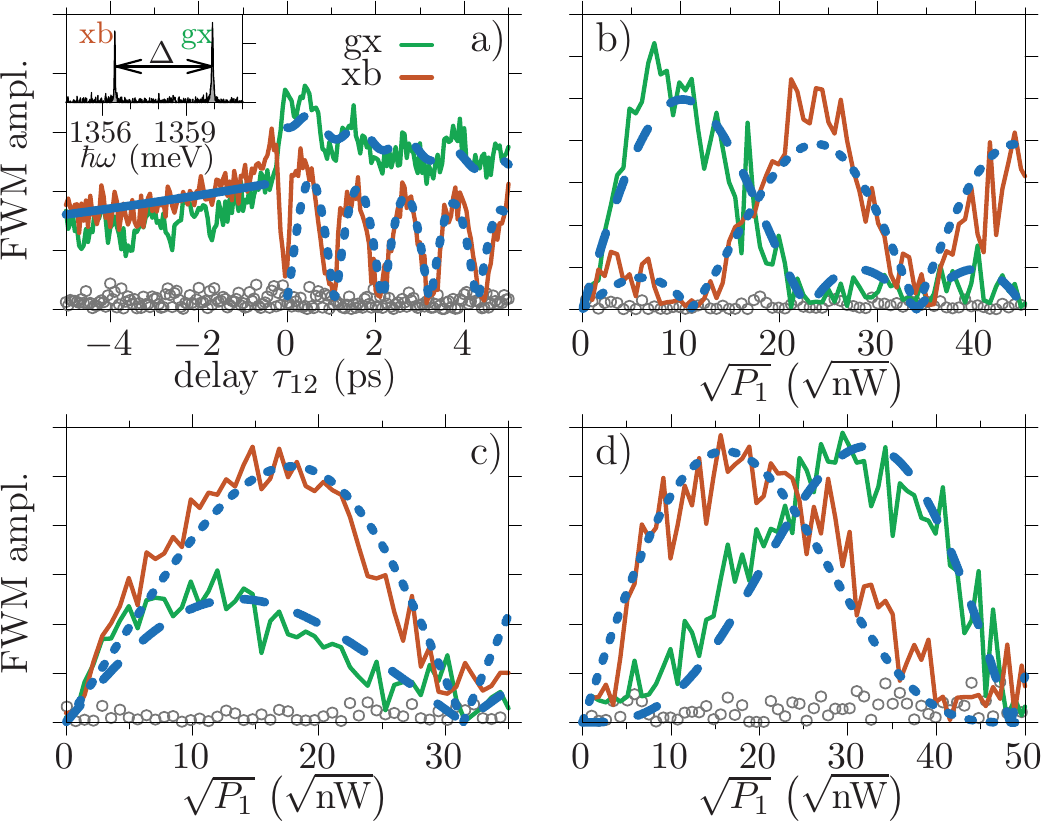}
\caption{
{\bf Rabi rotations on an exciton-biexciton system.}
a)~gxb system identified by measuring its coherence dynamics via the $\tau_{12}$-dependence of the FWM signals under co-linear polarization of $\Ed$. gx (xb) is represented with green (orange) traces, open symbols indicate the noise level. Fitted curves in blue. Inset: FWM spectrum at $\tau_{12}=0.5$~ps. b,c,d) Measured and calculated FWM amplitudes as a function of $\theta_1\sim \sqrt{P_1}$ for fixed $\theta_2$, $P_2=(250,\,600,\,1300)$~nW, respectively. The QD is the same as in Fig.~\ref{fig:2level_exp}.
\label{fig:3level_exp}
}
\end{figure}

The intensity dependence of the FWM signals becomes more intriguing when going beyond a two-level system. The natural extension is to consider the linearly polarized three-level system gxb in a neutral QD, as described in Sec.~\ref{sec:theory}. Such a gxb-system, with a BBE of $\Delta=3.6$~meV, has been identified by measuring the $\tau_{12}$-dependence of the FWM signals, as described in Ref.~[\onlinecite{MermillodOptica16}]. It was carried out on the same QD as in Fig.~\ref{fig:2level_exp}, but now under co-linear polarization aligned along one of the fine-structure axis. The characterization of the system is presented in Fig.~\ref{fig:3level_exp}~a). The measured coherence dynamics of the gx- and the xb-transition display pronounced beatings for $\tau_{12}>0$, with a period of $T_{\Delta}=2\pi\hbar/\Delta$. Additionally, we measure FWM signals for $\tau_{12}<0$, equally strong on both transitions, induced by the two-photon coherence between the ground state $\ket{g}$ and the biexciton $\ket{b}$.\cite{MermillodOptica16} The FWM spectrum of the system is shown in the inset, which confirms the BBE between the two transitions.

\begin{table}
\caption{\label{tab:parameters} Parameters for the pulses used in Fig.~\ref{fig:3level_exp}.}
\begin{ruledtabular}
\begin{tabular}{cccc}
panel & $\sqrt{P_2}$ \big($\sqrt{\rm nW}$\,\big) & $\sqrt{2}\,\theta_2^{\rm exp}$ ($\pi$) & $\theta_2^{\rm fit}$ ($\pi$)\\
\hline
b) & $15.8$ & $0.85$ & $0.7$\\
c) & $24.5$ & $1.35$ & $1.25$\\
d) & $36.1$ & $1.95$ & $2$\\
\end{tabular}
\end{ruledtabular}
\end{table}

In Fig.~\ref{fig:3level_exp}~b)-d) we present the measured FWM amplitudes of gx (green) and xb (orange) and the corresponding simulations (blue) as a function of $\sqrt{P_1}$ ($\theta_1$) for fixed $\sqrt{P_2}$ ($\theta_2$), respectively. Note that only in the $\chi^{(3)}$ regime, i.e., for small pulse areas, we recover that the gx signal is twice as strong as the xb one, as expected when considering Feynman diagrams of the gxb system.\cite{KasprzakNJP13} With increasing $\sqrt{P_1}$ ($\theta_1$) the FWM amplitudes exhibit oscillations on both, the gx- and the xb-transition. But now the generated Rabi rotations depend strongly on $\sqrt{P_2}$ ($\theta_2$) and are more complex than in the case of a two-level system in Fig.~\ref{fig:2level_exp}. Comparing the measured data with the simulated curves we find an excellent agreement. For the simulations we choose $\tau_{12}=0$ and added an exponential decay for the $\theta_1$-dependence in Eqs. \eqref{eq:FWM_3} as explained in Sec.~\ref{sec:exp_2}. To give a quantitative comparison between the pulse intensities in the experiment and the pulse areas in the theory, we summarize the crucial parameters in Tab.~\ref{tab:parameters}. From the pulse intensities $P_2$ in each panel [b)-d)] in Fig.~\ref{fig:3level_exp}, listed in the table, and the calibration of the pulse areas in the experiment in Eq.~\eqref{eq:calibration} we find the pulse areas in the experiment $\theta_2^{\rm exp}$. Due to the rescaling of the transition matrix element in Eq.~\eqref{eq:M_3L} we have to compare the scaled value $\sqrt{2}\,\theta_2^{\rm exp}$ with the theoretical fitted pulse areas $\theta_2^{\rm fit}$ in the table. Through the three values at hand, we find a reasonable agreement between experiment and theory.

\begin{figure}
\includegraphics[width=\columnwidth]{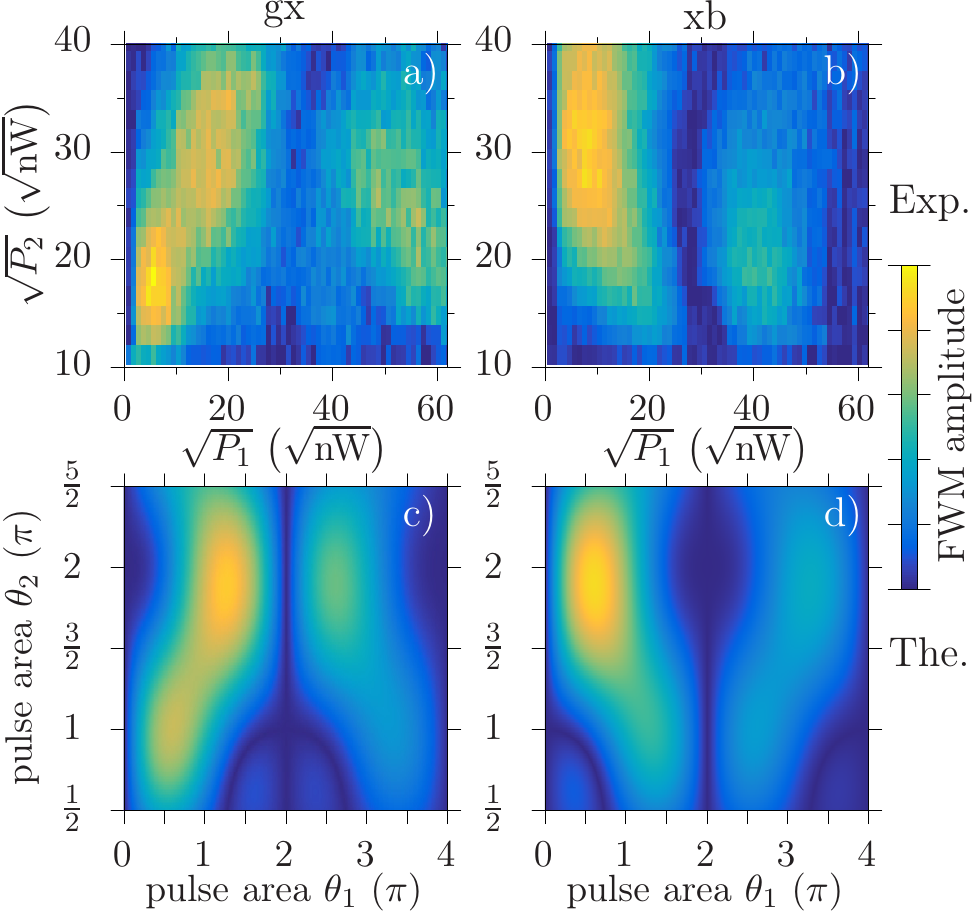}
\caption{
{\bf 2D Rabi rotations on an exciton-biexciton system.}
a,b)~Measured $(\sqrt{P_1},\sqrt{P_2})$-dependence of the FWM amplitude at gx in a) and xb in b). c,d) Corresponding simulations with additional exponential damping.\label{fig:2D_exp}
}
\end{figure}

We want to expand the measurements to a continuous picture, as it was given in Fig.~\ref{fig:3level}~c,d). This illustrates the interplay between the amplitudes of the gx- and the xb-transition. The result of the measurement is shown in Fig.~\ref{fig:2D_exp}~a) and b) for the gx and the xb FWM signal, respectively, as a function of $\sqrt{P_1}$ and $\sqrt{P_2}$. To achieve a reasonable agreement of the simulation with the measured data, we add an exponential damping to both pulse areas and show the same 2D Rabi rotations as in Sec.~\ref{sec:theory} but with adjusted axis ranges and decay rate in Fig.~\ref{fig:2D_exp}~c) and d) corresponding to a) and b). We directly see the excellent agreement of the pictures. The most striking feature is that both signals vanish for $\theta_1\approx 2\pi$. For smaller $\theta_1$ the gx-signal maximum in a) follows a positive slope, while the xb-signal maximum in b) tends in the other direction, i.e., to smaller $\theta_1$ for growing $\theta_2$. These signatures are also found in the simulations in c) and d). Although the signals are much weaker for $\theta_2>2\pi$, which makes an identification of the patterns quite difficult, we see that the maximum for gx in a) has a negative slope as in the simulation in c). The maximum in b) for large $\theta_1$ shifts to larger values for a growing $\theta_2$, which is also in agreement with the theory in d).

\begin{figure}
\includegraphics[width=0.75\columnwidth]{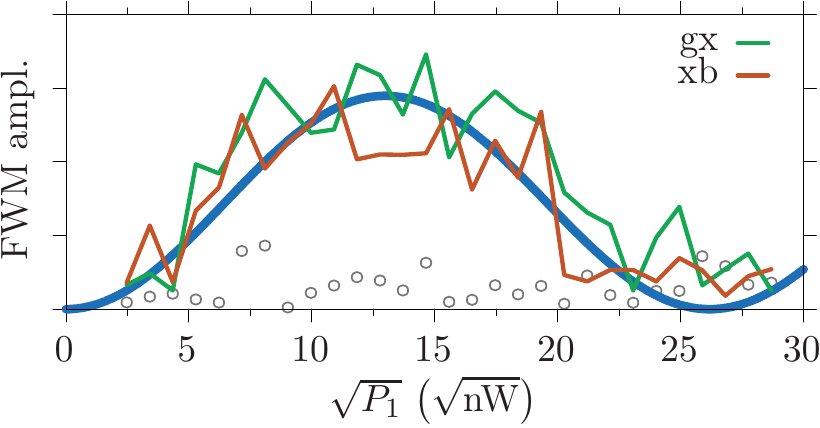}
\caption{{\bf FWM for two-photon coherence.}
FWM amplitudes as a function of pulse intensity $\sqrt{P_1}$ ($P_2=600$~nW) for a negative delay of $\tau_{12}=-2$~ps. gx (xb) is represented with a green (orange) trace, open symbols indicate the noise level. Fitted curve in blue.  
\label{fig:TPC}
}
\end{figure}

Finally, we want to focus on the Rabi rotations of the two-photon coherence measurement, i.e., for negative delays. The two FWM signals are shown in Fig.~\ref{fig:TPC} as a function of the amplitude of the first pulse $\sqrt{P_1}$. Both signals agree very well, as it is expected from the theory. The theoretical fit following Eq.~\eqref{eq:TPC} is given as blue line. We find a good agreement between the measured data and the predicted $\sin^2(\theta_1/2)$ dependence. Note that the data was taken on a different QD than the data for the rest of the section, which explains the slight discrepancy of the pulse areas. We here find that $\sqrt{P_1}\approx 25$~$\sqrt{\rm nW}$ corresponds to $\theta_1=2\pi$, while in Fig.~\ref{fig:2D_exp} $\theta_1=2\pi$ is found for $\sqrt{P_1}\approx 30$~$\sqrt{\rm nW}$.

\section{Phonon induced dephasing}
\label{sec:PID}
This last section revisits the exciton-phonon coupling in a two-level system. The aim is to demonstrate that the dynamics in such basic two-level systems does depend on the pulse areas when the coupling to phonons has a strong influence. The excitation of an exciton in a single QD on the ultrafast timescale is accompanied by the emission of a phonon wave packet.\cite{WiggerJPC14} The reason is the rapid deformation of the lattice in the QD region to form the new equilibrium state in the presence of the exciton, i.e., a polaron. If this polaron creation happens on the picosecond timescale or faster, a phonon wave packet is emitted as a shock wave. The phonon creation process leads to the loss of coherence in the excitonic part of the system, which can be detected as rapid drop of the FWM signal on the order of a few picoseconds, known as PID. In a recent study we have brought investigations of the PID effect from QD ensembles\cite{BorriPRL01, BorriPRB05} to a single QD.\cite{jakubczykACSP16} We found that the effect gets more pronounced when increasing the temperature. Calculations predict that the PID effect should not depend on the pulse areas  in the $\delta$-pulse limit. Interestingly, when extending the pulses into the range of a few hundred femtoseconds the coupling efficiency between the exciton and the phonons should depend strongly on the pulse intensity as explained above.\cite{WiggerJPC14} While in the $\delta$-pulse limit only single wave packet emission can be explained, extended excitations may lead to multiple creations and destructions of the polaron during a single pulse. These also induce the emission of sequences of multiple phonon wave packets.\cite{WiggerJPC14}

\begin{figure}
\includegraphics[width=\columnwidth]{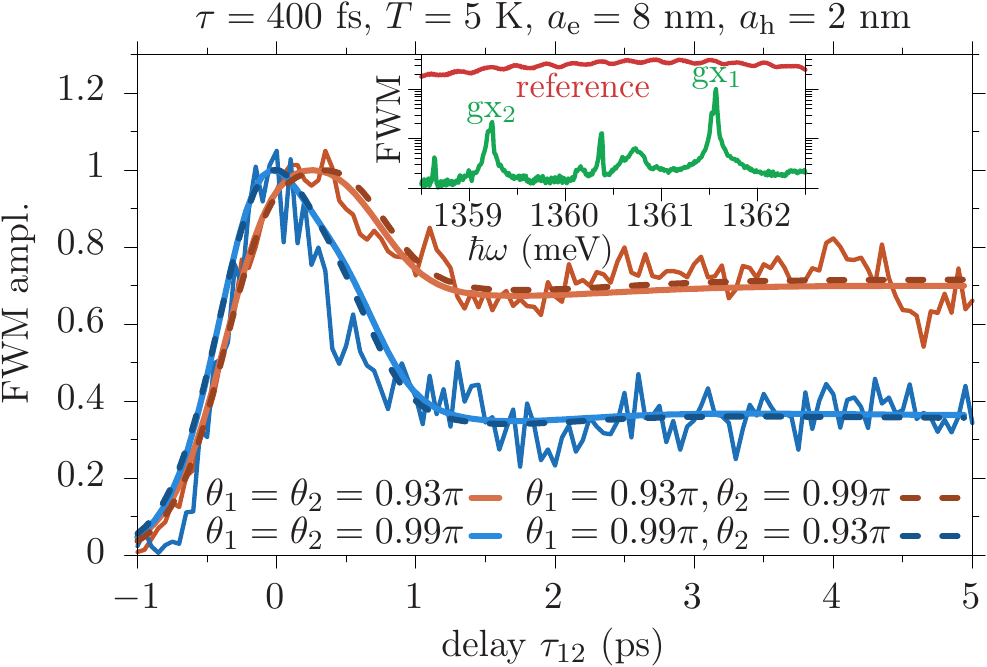}
\caption{
{\bf Phonon-induced dephasing.}
FWM amplitude as a function of delay $\tau_{12}$ for pulse durations of $\tau=400$~fs at $T=5$~K. The fitted electron and hole localization lengths are $a_{\rm e}=8$~nm and $a_{\rm h}=2$~nm, respectively. Inset: FWM spectrum in green, laser pulse spectrum in red.
\label{fig:PID}
}
\end{figure}

Figure~\ref{fig:PID} shows an exemplary result of the PID effect after the excitation with longer laser pulses. It shows the FWM amplitude as a function of the delay $\tau_{12}$ for gx-transitions in two different QDs but in the same defect. A FWM spectrum taken from this defect is shown as an inset in Fig.~\ref{fig:PID} in green together with the laser pulse spectrum in red. The blue data points are taken from the transition in the center of the pulse spectrum (gx$_1$ in the inset), the orange ones from a transition at the edge of the pulse spectrum (gx$_2$ in the inset). Thus the latter one was excited with less intensity. Obviously, the PID drop is much more pronounced for the system that is driven with more intensity. From simulations with extended laser pulses of $\tau=400$~fs within the correlation expansion formalism\cite{krugelPRB06} we found that the curves can be reproduced by choosing different pulse areas for the two driving pulses. The best agreements was achieved for $\theta_1=\theta_2=0.93\pi$ and $0.99\pi$, which are shown as solid orange and blue line in Fig.~\ref{fig:PID}, respectively. The difference of the pulse areas is reasonable for the two positions in the pulse spectrum in the inset. In a two-level system that is not coupled to phonons $\theta_1\approx\pi$ would lead to a vanishing FWM signal because the polarization after the first pulse would be zero. The phonons now lead to dephasing during the extended first pulse, which results in a finite polarization that can be transferred into the FWM signal. Note that the data was taken for a temperature of $5$~K. For such low temperatures the theory in the $\delta$-pulse limit does not predict drops of the FWM signal on the order of 60\%, as it is found here. Therefore the increased interaction efficiency between the exciton and the phonons for the longer pulse excitation is responsible for the strong PID effect.

This fit of the dynamics of the PID drop motivated us to use the localization lengths of $a_{\rm e}=8$~nm and \linebreak$a_{\rm h}=2$~nm for electron and hole, respectively, as they were used for the phonon spectral density in Fig.~\ref{fig:spec_dens}. These sizes mimic a rather flat lens shaped dot, which is in agreement with STEM measurements on similar QDs.\cite{braunPRB16}

To show that the PID effect is dominantly governed by the phonon emission after the first laser pulse, which launches the phonon wave packet that leads to the PID effect, we additionally show simulations for \linebreak$(\theta_1,\theta_2)=(0.93, 0.99)\pi$ and $(0.99, 0.93)\pi$ as dashed lines. The deviation from the $\theta_1=\theta_2$-cases are very small. Because both cases agree in the first pulse area this confirms the assumption. The second pulse just transfers the coherence into the FWM signal and its detailed influence on the coupled exciton-phonon system is not of great importance.

There is still one open aspect about the dynamics depending on the pulse area, i.e., the shift of the FWM maximum with changing pulse area mentioned in Sec.~\ref{sec:exp}. In Fig.~\ref{fig:PID} we see this effect quite clearly. Although here the pulses are longer than for the Rabi rotations investigated in Sec.~\ref{sec:exp}, this shift still remains for pulses in the $\tau\lesssim 200$~fs range, but is less pronounced.

\section{Conclusions}
In summary, we have studied Rabi rotations within different few-level systems of a QD exciton complex. We used natural photonic defects that allowed for a efficient in- and out-coupling of the optical fields, facilitating the retrieval of FWM signals. The Rabi rotations in a two-level system that were realized by circularly polarized driving, worked as calibration for the pulse areas in the experiment. They also showed that the details of the exciton-phonon coupling play an important role when describing the damping of the rotations with increasing pulse area. By choosing linearly polarized excitations we could investigate a three-level system. Here, the Rabi rotations were much more involved and a complex interplay between the two pulse areas was found. In a next step, we showed an example for the importance of the exact knowledge of the pulse areas. We pointed out that the PID effect can become very pronounced for pulse durations in the few hundred femtosecond range and that its strength decisively depends on the pulse areas in the experiment. For pulse areas that significantly exceed $\pi$ multiple phonon wave packet emissions take place, which should also lead to more complex dynamics in the FWM signal. A detailed investigation of the PID effect is still needed. The influence of pulse durations, amplitudes and temperature has to be studied in future projects.

\section*{Acknowledgement} We acknowledge the financial support by the
European Research Council (ERC) Starting Grant PICSEN (grant no. 306387). We thank S.~Pairis and J.-F.\,Motte for performing AFM and SEM + FIB measurements, using NanoFab facilities at Institut N\'{e}el, CNRS Grenoble. We further thank Sebastian L\"uker for support with the implementation of the phonon coupled model. JK thanks Wolfgang Langbein for continuous discussions and support.

%\bibliography{IOP,defects}


\end{document}